
%
%
\def\half {{1 \over 2}}
\def\dz{{\partial _z}}
\def\dzb{{\partial _{\bar z}}}
\def\Nm{{\nabla_{\mu}}}
\def\Nbm{{\bar\nabla_{\dot\mu}}}
\def\fm{{f_{\mu}^m}}
\def\fbm{{\bar f_{\dot\mu}^m}}
\def\Xm{{X^m}}
\def\Xbm{{\bar X^m}}
\def\Tm{{\Theta^\mu}}
\def\T{{\Theta}}
\def\To{{\Theta^1}}
\def\tone{{\theta^1}}
\def\Tbm{{\bar \Theta^{\dot\mu}}}
\def\Tbo{{\bar \Theta^1}}
\def\Tb{{\bar \Theta}}
\def\tbo{{\bar \theta^1}}
\def\hb{{\bar h}}
\def\k{{\kappa}}
\def\kb{{\bar\kappa}}
\def\Db{{\bar D}}
\def\gm {{\gamma^m_{\mu{\dot \mu}}}}
\def\TgT {{\Tm\gm\Tbm}}
\def\lo {{\lambda^1}}
\def\lbo {{\bar \lambda^1}}
\def\m {{\mu}}
\def\md {{\dot\mu}}

\def\dzxp {{\dz x^{3+0} +\half\tone\dz\tbo+\half\tbo\dz\tone}}

\def\wb {{\bar w}}
\def\dXm{{\partial _{\Xm}}}

\def\dTm{{\partial _{\Tm}}}
\def\dTbm{{\partial _{\Tbm}}}

\def\Km{{K_m}}
\def\Kmu{{K_{\mu}}}
\def\Kbmu{{\bar K_{\md}}}
\def\Kbm{{\bar K_{\bar m}}}
\def\KbM{{\bar K_{\bar M}}}

\tolerance=5000
\footline={\ifnum\pageno>1
       \hfil {\rm \folio} \hfil
    \else \hfil \fi}

\overfullrule=0pt 
\baselineskip=18pt
\raggedbottom
\centerline{\bf A New Sigma Model Action for the}
\centerline{\bf Four-Dimensional Green-Schwarz Heterotic Superstring}
\vskip 24pt
\centerline{Nathan Berkovits}
\vskip 24pt
\centerline{Math Dept., King's College, Strand, London, WC2R 2LS, United
Kingdom}
\vskip 12pt
\centerline{e-mail: udah101@oak.cc.kcl.ac.uk}
\vskip 12pt
\centerline {KCL-TH-93-3}
\vskip 12pt
\centerline {February 1993}
\vskip 96pt
\centerline {\bf Abstract}
\vskip 12pt

The sigma model action described in this paper
differs in four important features from the usual sigma model
action for the four-dimensional Green-Schwarz heterotic superstring
in a massless background. Firstly, the action is constructed on an
N=(2,0) super-worldsheet using a Kahler potential and an
Ogievetsky-Sokatchev constraint; secondly, the target-space background
fields are unconstrained; thirdly, the target-space dilaton couples
to the two-dimensional curvature; and fourthly, the action reduces
in a flat background to a free-field action.

A conjecture is made for generalizing this N=(2,0) sigma model action
to the ten-dimensional Green-Schwarz heterotic superstring in a manner
that preserves these four new features.
\vfil
\eject
\centerline{\bf I. Introduction}
\vskip 12pt

The construction of two-dimensional sigma model actions for the
bosonic and Neveu-Schwarz-Ramond strings in massless backgrounds
has been a useful tool for determining string-corrected equations
of motion for the massless fields.$^1$ These equations of motion are
obtained by calculating the $\beta$-functions of the non-linear
sigma model perturbatively in $\alpha '$ and demanding that
they vanish. Although in principle, the string-corrected equations
of motion can also be determined from the S-matrix scattering
amplitudes, the $\beta$-function technique has proven to be
much more effective.

However in order for the perturbative expansion in $\alpha '$ of the
$\beta$-function to make sense, the sigma model action must become
a two-dimensional free-field action when $\alpha '=0$. By using a
normal coordinate expansion for the background fields,$^2$ this
condition implies that the sigma model action in a flat background
must also be a free-field action. Note that such a free-field
condition is also needed for calculating S-matrix scattering amplitudes.

For the case of the Green-Schwarz superstring, the usual sigma model
action$^3$ does not satisfy this free-field condition. One attempt at
avoiding this problem is to classically gauge-fix all of the fermionic
Siegel symmetries to semi-light-cone gauge$^4$ where $\gamma^+ \theta =0$,
and then rescale $\gamma^- \theta$ to $\sqrt{\dz x^+} (\gamma^- \theta)$.
Although the resulting sigma model action in a flat background is
a free-field action, it contains several strange features.$^{5,6}$

Firstly, because the semi-light-cone gauge choice is not possible$^7$
at points where $\dz x^+ =0$, the target-space background fields
must satisfy the light-cone gauge condition$^8$ that their + components
vanish and that they are independent of $x^-$ (this condition allows
$x^+$ to have classical solutions with $\dz x^+$ nowhere vanishing).\footnote
\dag{ The authors of reference 5 disagree that this light-cone gauge
condition is necessary, and instead impose only the restriction
that $R_{abc+}=R_{abc-}=0$.}
Secondly, the requirement that the original sigma model action is
classically invariant under Siegel transformations imposes certain
torsion constraints on the background fields. (In ten dimensions, these
torsion constraints force the background fields on-shell,$^9$ so vanishing
of the one-loop $\beta$-function does not further restrict them.$^5$
In four dimensions, these torsion constraints do not force the
background fields on-shell,$^{10}$ however vanishing of the one-loop
$\beta$-function is not enough to determine their equations of motion.$^6$)
Thirdly, the target-space dilaton field does not couple to the
two-dimensional curvature,$^{11}$ so one does not get the usual relationship
between the string coupling constant and the dilaton zero mode. And
fourthly, the conformal anomaly of this free-field action is
$c=d+\half (d-2) -26$, which is non-zero even when $d=10$.$^{12}$

An alternative approach to the sigma model action for the Green-Schwarz
heterotic superstring was developed more recently using two-dimensional
super-worldsheets.$^{13-17}$ After imposing an STVZ-like constraint$^{18}$
on the
target-space coordinates, sigma model actions for the Green-Schwarz
heterotic superstring  have been constructed with N=(1,0), (2,0), (4,0),
and (8,0) worldsheet supersymmetry.$^{15,17}$ Although all of these actions
coincide classically with the usual Green-Schwarz sigma model action,
only the N=(2,0) action$^{15}$ has been shown to reduce in a flat background
to a free-field action which can be consistently quantized in ten
dimensions.$^{19,20}$ The disadvantage of these actions is that the STVZ-like
constraint requires the background fields to satisfy the same torsion
constraints as in the usual Green-Schwarz sigma model, and the
dilaton field still does not couple to the two-dimensional curvature.

In this paper, a new sigma model action with N=(2,0) worldsheet
supersymmetry is constructed for the Green-Schwarz heterotic superstring.
After replacing the STVZ-like constraint with the Ogievetsky-Sokatchev
constraint, this sigma model action is constructed out of a Kahler
potential with no extra constraints on the background fields and
with the standard Fradkin-Tseytlin coupling of the target-space
dilaton field to the two-dimensional curvature. Although this N=(2,0)
action can be constructed for four, six, and ten-dimensional target spaces,
only the four-dimensional case will be investigated in detail and
shown to describe a background of minimal d=4 N=1
supergravity and super-Yang-Mills coupled to an antisymmetric
tensor field. It is conjectured in the conclusion of this paper that
the analogous N=(2,0) sigma model action for a ten-dimensional
target space describes a background of d=10 N=1 supergravity and
super-Yang-Mills.

\vskip 12pt
\centerline{\bf II. The Sigma Model Action in a Flat Target Space}
\vskip 12 pt

In four dimensions, the new sigma model action is a curved target-space
generalization of the N=(2,0) twistor-string action first described
by Ivanov and Kapustnikov.$^{16}$
This action in a flat background was constructed
using a four-dimensional complex target space, $[\Xm, \Xbm, \Tm, \Tbm]$ where
$m=0$ to 3 and $\m ,\md$=1 to 2,
together with the reality condition,
$$\Xm -\Xbm =i\TgT.\eqno(II.1)$$

As two-dimensional superfields on the N=(2,0) super-worldsheet, these
target-space coordinates satisfy the chirality conditions,
$$\Db \Xm= \Db\Tm = D\Xbm = D\Tbm =0,  \eqno(II.2)$$
where $D=\partial _\k +{i \over 2}\kb\dz$,
$\Db=\partial _\kb +{i\over 2}\k\dz$,
and the two-dimensional Minkowski-space super-worldsheet is parameterized
by the coordinates $[z,\k,\kb; \bar z]$ with $\k$ the complex conjugate
of $\kb$, but $z$ and $\bar z$ independent real variables.

In the presence of these reality and chirality constraints, the flat
target-space action of Ivanov and Kapustnikov is:
$$S=i\int d^2 z d^2 \k [ X_m \dzb \Xbm +i(X_m\Tbm\gm)\dzb\Tm -i(\bar X_m\Tm\gm)
\dzb\Tbm],\eqno(II.3)$$
which takes the following form after using the reality constraints to
express all superfields in terms of $X^{1+i2}$, $\bar X^{1-i2}$,
$X^{3-0}$, $\bar X^{3-0}$, $\Theta^1$, and $\bar\Theta^1$ (e.g.,
$X^{1-i2}=\bar X^{1-i2} +i\bar\Theta^1 \Theta^2$,
$\Theta^2=i\Db \bar X^{1-i2}/ \Db\bar \Theta^1$, etc.):$^{19}$
$$S=i\int d^2 z d^2 \k [ X^{1+i2} \dzb \bar X^{1-i2}
+i\bar W\dzb \Theta^1 -i W\dzb \bar \Theta^1] ,\eqno(II.4)$$
where $W= \bar X^{3-0}\Theta^1$ and $\bar W= X^{3-0}\bar\Theta^1$.

The only remaining constraints implied by equation (II.1) are:
$$D\Theta^1 \bar D\bar \Theta^1
-{i\over 2}
(\To\dz\Tbo+\Tbo\dz\To)=\half\dz(X^{3+0}+\bar X^{3+0}) \qquad {and}\qquad
-i\Theta^2\bar\Theta^2= X^{3-0}- \bar X^{3-0}.\eqno(II.5)$$
The first of these constraints can be solved by bosonizing
the components of
$\To=\tone +\k \lo$,
$ \Tbo=\tbo +\kb \lbo$,
$DW=w+\kb\varepsilon$, and
$\bar D\bar W=\bar w+\k\bar\varepsilon$
in the following way:$^{20}$
$$\lo=(\dzxp)
 e^h +e^{-\hb},\quad \lbo=e^{-h}, \eqno(II.6)$$
$$w=e^h (\dz h+\dz\hb+x^{3-0} (\dzxp))+x^{3-0} e^{-\hb},\quad
 \wb=x^{3-0} e^{-h},$$
where $x^{3\pm 0}$ is the
lowest component of $\half(X^{3\pm 0}+\bar X^{3\pm 0})$ and
$h,\bar h$ are chiral bosons with screening charge $-1$ that
satisfy $h(y) \hb (z) \to \log (y-z)$ as $y\to z$.
Because the operator-products of $\lo$, $\lbo$, $w$, and $\bar w$,
are just free-field operator products, the action of equation (II.4)
remains a free-field action when expressed in terms of $X^{1+i2}$,
$\bar X^{1-i2}$, $x^{3+0}$, $x^{3-0}$, $h$, $\hb$,
$\tone$, $\tbo$, $\hat\varepsilon$, and $\hat{\bar\varepsilon}$
($\hat\varepsilon\equiv\varepsilon
-\half\dz x^{3-0}\tone-x^{3-0}\dz\tone$ has no singularities near $x^{3+0}$).

The second constraint of equation (II.5) can be rewritten,
$$DX^{1+i2} \Db \bar X^{1-i2}+iDW \Db\Tbo -i\Db \bar W D\To=0,\eqno(II.7)$$
which is just the N=(2,0) stress-energy tensor for the free-field
action of equation (II.4). The conformal anomaly of this stress-energy
tensor in $d$
dimensions is $c=(d-2)+\half (d-2)+2(-2-1)$, which cancels the conformal
anomaly contribution of the N=(2,0) ghosts ($c=-6$) when $d=10$.$^{19}$

Although manifest SO(3,1) super-Poincar\'e invariance has been
broken, this free-field action and stress-energy tensor can
easily be checked to be Lorentz invariant by explicitly constructing
the SO(3,1) super-Poincar\'e generators out of the free fields. For a
discussion of this procedure, see section III of reference 21 where
these generators were explicitly constructed for the ten-dimensional
case.
\vskip 12pt
\centerline{\bf III. The Massless Background Fields}
\vskip 12 pt

The first step in generalizing to a curved target space is to replace
the reality constraint of equation (II.1) with the constraint,
$$\Xm -\Xbm= iH^m ( X+\bar X, \Theta, \bar \Theta),$$
where $H^m$ is a real superfield. This reality constraint first
appeared in the work of Ogievetsky and Sokatchev in their description
of conformal supergravity,$^{22}$ and more recently was used by Delduc
and Sokatchev to write an N=2 worldline supersymmetric action for the
four-dimensional superparticle in a supergravity and super-Yang-Mills
background.$^{23}$

It is convenient to define fermionic target-space derivatives,
$$\Nm=\partial_{\Tm} +i \fm \partial_{\Xm}\qquad {and}\qquad
\Nbm=\partial_{\Tbm} +i \fbm \partial_{\Xbm},\eqno(III.1)$$
where $\fm=A_n^m \dTm H^n $,
$\fbm=-\bar A_n^m \dTbm H^n $, and $A_n^m$ is the inverse matrix of
$(\delta_m^n -i\dXm H^n)$,
$\bar A_n^m$ is the inverse matrix of
$(\delta_m^n +i\dXm H^n)$. Note that $\fm$ and $\fbm$ are uniquely
determined by requiring that
$$[\Nm ,\,\Xm-\Xbm-iH^m]=
[\Nbm ,\,\Xm-\Xbm-iH^m]=0.\eqno(III.2)$$

Under the holomorphic coordinate transformations, $X^m\to X'^m (X, \Theta)$
and $\Tm\to \Theta'^{\mu} (X,\Theta)$, it is easy to check that
$\Nm\to \nabla'_{\mu}= A_{\mu}^{\nu} \nabla_{\nu}$, where
$H^m\to H'^m=H^m -i(X'^m-\bar X'^m)+i(X^m-\bar X^m)$ and $A_{\mu}^{\nu}$
is the inverse matrix of $\nabla_{\nu}\Theta'^{\mu}$.

The non-gauge degrees of freedom in the Ogievetsky-Sokatchev superfield,
$H^m$, can be determined by using some of these holomorphic
coordinate transformations to gauge-fix to the form:
$$H^m=\Tm\Tbm e^m_{\m\md} + \Tm (\Tb)^2 \xi^m_{\mu}
+ \Tbm (\T)^2 \bar\xi^m_\md + (\T)^2(\Tb)^2 a^m.\eqno(III.3)$$
The remaining holomorphic coordinate transformations can be used to
further gauge away 11 components of $e^m_{\m \md}$ (4 coordinate +
6 Lorentz + 1 scale), 8 components of $\xi^m_\m$ and $\bar\xi^m_\md$
(4 Q-supersymmetries and 4 S-supersymmetries), and one component
of $a^m$ (1 chiral), leaving the usual conformal supergravity
multiplet of 8 bosons and 8 fermions.$^{22}$

The next step in constructing an N=(2,0) sigma model action is to
introduce a Kahler potential on the complex
four-dimensional target space, $K_M (X,\bar X,\T,\Tb)$, and its
compex conjugate, $\KbM (X,\bar X,\T,\Tb)$, where $M$ takes four
complex bosonic values ($m=0$ to 3) and two complex
fermionic values ($\m$=1 to 2).
As usual in Kahler geometry,$^{24}$ the Kahler metric and torsion potential
will be defined in terms of $K_M$ by:
$$G_{M \bar N}=\partial_M \bar K_{\bar N} +(-1)^{s(M)s(\bar N)}
\partial_{\bar N} K_M\qquad {and} \qquad
B_{M \bar N}=\partial_M \bar K_{\bar N} -(-1)^{s(M)s(\bar N)}
\partial_{\bar N} K_M,\eqno(III.4)$$
where $s(M)=0$ (or 1) if M takes bosonic (or fermionic) values.

Because $K_M$ and $H^m$ both describe the target-space geometry,
they are not independent superfields. The constraint that relates them
is:
$$h_{\m \md}\equiv G_{\m \md}+i\fm G_{m \md} -i\fbm G_{\m\bar m} -
\fm \bar f_{\md}^n G_{m \bar n}=0,\eqno(III.5)$$
where $G_{M \bar N}$ and $\fm ,\fbm$ are defined in equations (III.4) and
(III.1). Under holomorphic coordinate transformations, $h_{\m \md}\to
h'_{\m \md}=A_\m^\nu \bar A_\md^{\dot\nu} h_{\nu \dot\nu}$ where
$A_\m^\nu$
is the inverse matrix of $\nabla_{\nu}\Theta'^{\mu}$, so the constraint is
coordinate-independent.

Under the following transformations of $K_M$, $\delta G_{M \bar N}=0$
and $\delta B_{M \bar N}=\partial_M \partial_{\bar N} F$ (which leaves
the field strength $\partial_P B_{M \bar N}
-(-1)^{s(M)s(P)}
\partial_M B_{P \bar N}$ unchanged):
$$\delta K_M = \Lambda_M (X,\T)+i\partial_M F(X,\bar X,\T,\Tb),\qquad
\delta \KbM = \bar\Lambda_{\bar M} (\bar X,\Tb)
-i\partial_{\bar M} F(X,\bar X,\T,\Tb),\eqno(III.6)$$
where $\Lambda_M$ is holomorphic and $F$ is real.

Using the $\Lambda_\mu$ and $\Lambda_m$ transformations, it is
possible to gauge-fix $\hat K_\m\equiv K_\m +i\fm K_m$ to the
following form:
$$\hat K_\m=  (\Tb)^2 \phi_\m +\Theta^\nu(\Tb)^2 t_{\m \nu}
+ i(\T)^2 (\Tb)^2 \psi_\m.\eqno(III.7)$$
Furthermore, choosing $F= \Tm\Tbm c_{\m \md} +\Tm (\Tb)^2 c_\m
+ \Tbm (\T)^2 \bar c_\md+ (\T)^2 (\Tb)^2 c+i(X^m-\Xbm -iH^m)C_m(\T,\Tb)$,
one can gauge away
$t_{\m\nu} \epsilon^{\m\nu}-
\bar t_{\md\dot \nu} \epsilon^{\md\dot\nu}$, three components of the
antisymmetric tensor field, $b_{mn}=\gamma_m^{\m\md}\gamma_n^{\nu\dot\nu}
((t_{\mu\nu}+t_{\nu\mu})\epsilon_{\md\dot\nu}+
(\bar t_{\md\dot\nu}+\bar t_{\dot\nu\md})\epsilon_{\m\nu}))$, all of
$\phi_\m$ and $\bar\phi_\md$, and the complete superfield $K_m +
\bar K_{\bar m}$ (note that $\hat K_\m$ and $\hat\Kbmu$
are unaffected by the gauge
transformation involving the real superfield $C_m$). Since $0=h_{\m \md}=
-i\bar K_{\bar m}\Nm \fbm
+iK_{m}\Nbm \fm
+\Nm \hat \Kbmu -\Nbm \hat \Kmu$, the remaining components of $K_m$
and $\Kbm$ are determined from $\hat\Kmu$,
$\hat \Kbmu$, and $H^m$, and therefore the
only non-gauge degrees of freedom in $K_M$ are a real scalar, $\sigma
\equiv t_{\m \nu}\epsilon^{\mu\nu}+\bar t_{\md\dot\nu}\epsilon^{\md
\dot\nu}$, a gauge-fixed antisymmetric tensor, $b_{mn}$, and complex
spinors, $\psi_\m$ and $\bar\psi_\md$, totalling four bosons and
four fermions.

Although the usual interpretation of the $\sigma$, $\psi_\m$, and
$\bar\psi_\md$
fields is as the target-space dilaton and dilatinos, it is more
natural to identify them as the determinant of the vierbein,
$\det (e_{\m \md}^m)$, and the gamma-matrix traces of the gravitinos,
$\gamma_m^{\m \md}\xi_\m^m$ and
$\gamma_m^{\m \md}\bar\xi_\md^m$. This is because using the holomorphic
coordinate transformations that give rise to scale and S-supersymmetries,
$\Tm\to a\Tm +i(\T)^2\alpha^\m$, one could have gauge-fixed these
$\sigma$ and $\psi_\m$ components of $\hat K_\m$ instead of gauge-fixing
the vierbein and gravitino components of the $H^m$ superfield.$^{10}$

With this identification of $\sigma$ and $\psi_\m$, one now needs
a superfield containing the target-space dilaton and dilatinos. This
can be accomplished with a complex holomorphic
scalar bosonic superfield,
$\Phi (X, \T)$, and its
complex conjugate, $\bar\Phi (\bar X,\Tb)$. Expanding in components,
$\Phi =\phi + i\Tm\chi_\m +i(\T)^2\rho$, where $\phi+\bar\phi$ is
the dilaton, $\chi_\m$ and $\bar\chi_\md$ are the dilatinos, and
$i(\phi-\bar\phi), \rho,\bar\rho$ are bosonic auxiliary fields. So the
total non-gauge degrees of freedom in $H^m$, $K_M$, and $\Phi$
are 16 bosons and 16 fermions, as in the usual minimal formulation
of N=1 Poincar\'e supergravity coupled to an antisymmetric tensor
field.

Finally, coupling the superstring to a super-Yang-Mills background
can be accomplished in the same way as for the superparticle,$^{23}$ by
introducing a real scalar superfield, $V^I(X,\bar X,\T,\Tb)$,
with the gauge invariance $\delta V^I=\Lambda^I(X,\T) +
\bar\Lambda^I(\bar X,\Tb)$ where $\Lambda^I$ is a holomorphic superfield
and $I$ labels the group generators. Since $V^I$ can be gauge-fixed
to the form, $V^I= \Tm\Tbm v_{\m \md}^I + \Tm (\Tb)^2 w_\m^I
+\Tbm (\T)^2 \bar w_\md^I +(\T)^2 (\Tb)^2 y^I$, where $v^I_{\m \md}$
contains the further gauge invariance, $\delta v^I_{\m \md}=
\gm \partial_m \lambda^I$, $V^I$ contains the usual non-gauge
degrees of freedom of 4$g$ bosons and
4$g$ fermions of N=1 super-Yang-Mills ($g$ is the dimension of the group).

\vskip 12pt
\centerline{\bf IV. The Sigma Model Action in a Curved Target Space}
\vskip 12 pt

After putting the two-dimensional super-vierbein into superconformal
gauge such that $$D_\k =  e^{\bar L(z-{i \over 2}\k\kb, \kb;\bar z)}
 (\partial_\k
+{i\over 2}\kb\dz) \qquad {and} \qquad
\bar D_\kb = e^{L(z+{i \over 2}\k\kb, \k;\bar z)}
 (\partial_\kb
+{i\over 2}\k\dz),\eqno(IV.1)$$
the sigma model action for the four-dimensional
Green-Schwarz heterotic superstring in a massless background is:
$$S=i\int d^2 z d^2 \k
[\Km \dzb X^m +\Kmu \dzb \Tm -\Kbm \dzb \Xbm -\Kbmu \dzb\Tbm +
iV^I j_I + \k\kb k  +\alpha '(\Phi\dzb\bar L -\bar\Phi\dzb L)],\eqno(IV.2)$$
where $\Xm -\Xbm =iH^m$, $H^m$ is determined by $K_M$ and $\KbM$
from the constraint $h_{\m \md} =0$ of equation (III.5), $j_I$ is any
right-moving current satisfying the commutation relations
$[j_I, j_J]= f^K_{IJ} j_K$, $k$ is the kinetic energy term for the
two-dimensional right-moving
fields in $j_I$, and $L,\bar L$ are chiral and anti-chiral
two-dimensional superfields containing the N=(2,0) superconformal
degrees of freedom. Note that the left-moving stress-energy
tensor is $\bar D_\kb \Km D_\k \Xm
+\bar D_\kb \Kmu D_\k \Tm
-D_\k \Kbm \bar D_\kb \Xbm
-D_\k \Kbmu \bar D_\kb \Tbm=-h_{\m \md}
D_\k \Tm \bar D_\kb \Tbm=0$. In a flat background,
$K_m=-\half\bar X_m$ and $\Kmu=i X_m \gm \Tbm$, so the lowest
component of $-i\bar D_\kb K_\m$
and $-iD_\k \Kbmu$
can be interpreted as a curved-space generalization of the twistor
fields,$^{25}$ $\bar w_\m =x_m \gm \bar\lambda^\md$
and $w_\md =\bar x_m \gm \lambda^\m$.

By performing the integration over $\k$ and $\kb$, one obtains
the following action:
$$S=\int d^2 z
[\,\Omega\, \eta_{ab} E^a_M E^b_N \,\dz y^M \dzb y^N
+b_{MN}\, \dz y^M \dzb y^N + A_M^I \, j_I \dz y^M +k\eqno(IV.3)$$
$$ +\alpha ' ((\Phi+\bar\Phi)\, R
+i(\Phi-\bar \Phi)\, F  +  i(\Nm\Phi) \,\lambda^\m \zeta
+i(\Nbm\bar\Phi)\, \bar\lambda^\md\bar\zeta)],$$
where
$M$ and $N$ range over four real bosonic values ($m,n=0$ to 3)
and four real fermionic values ($\mu,\nu=1$ to 2, $\dot\mu ,
\dot\nu$=1 to 2), $y^m=x^m$, $y^\m =\theta^\m$,
$y^\md =\bar\theta ^\md$, $E^A_M$ is the inverse super-vierbein obtained
from the covariant fermionic derivatives $\Nm$ and $\Nbm$ of equation
(III.1) (e.g., $E_\alpha^\mu=\delta_\alpha^\mu$ and $E_\alpha^m=
f_\alpha^m)$,
$b_{MN}$ and its field strength $\Omega$ are obtained from the
chiral spinor prepotential $\Sigma_\mu=(\bar\nabla)^2 \hat K_\mu$ (e.g.,
$b_{m\m}=\gamma_{m\, \m \md}
\bar\Sigma_{\dot\nu} \epsilon^{\md \dot\nu}$ and
$\Omega=\nabla_\mu \Sigma_\nu \epsilon^{\mu\nu}+
\bar\nabla_\md \bar\Sigma_{\dot\nu} \epsilon^{\md\dot\nu}$),
$A_M$ is obtained from the gauge-covariant derivatives
$e^{-V} \Nm e^V$ and $\Nbm$,$^{26}$ $R$ is the two-dimensional curvature,
$F$ is the field strength of the two-dimensional gauge field, $\zeta$
and $\bar\zeta$ are the field strengths of the two-dimensional
gravitinos, and $\lambda^\mu$ is the lowest component of $D_\k \Tm$.

The first part of this action is equivalent to the usual
four-dimensional Green-Schwarz sigma model action, except for the
new torsion constraints
on $E_A^M$ that $T_{a \alpha}^\beta=T_{a \alpha}^{\dot\beta}=0$
(as was shown in reference 10, this ``superconformal" Green-Schwarz
action is invariant under Siegel transformations).

The second part of the action is fundamentally different from the
usual Green-Schwarz action in that it contains couplings to the
N=(2,0) geometry. Note that not only the genus coupling constant,
but also the instanton theta parameter, can now be absorbed into the
zero modes of background fields.

\vskip 12pt
\centerline{\bf V. Conclusion}
\vskip 12pt

As discussed in the introduction, the usual Green-Schwarz sigma
model action has many unpleasant features that are not present in
the action of equation (IV.2). Because this new action reduces to a
free-field action in a flat background, it should be possible
to calculate the
$\beta$-function perturbatively and find string-corrected equations
of motion for the d=4 N=1 supergravity fields.

One possible generalization would be to construct a similar sigma model
action for the four-dimensional Green-Schwarz closed superstring
in an N=2 supergravity background. It seems likely that one should
use an N=(2,2) super-worldsheet and introduce a Kahler potential
which depends on complex target-space coordinates which are either chiral
or twisted-chiral
two-dimensional superfields,$^{27}$
i.e., are chiral or anti-chiral independently
in the left and right-handed directions.

A more interesting generalization would be to construct an N=(2,0)
sigma model
action for the ten-dimensional Green-Schwarz heterotic superstring.
An obvious guess for this action is to simply replace the four-component
complex vectors and two-component complex spinors with ten-component
complex vectors and sixteen-component complex spinors. The constraint
$h_{\m \nu}=0$ of equation (III.5) now contains 256 components which
restrict not only the imaginary part of $X^m$, but also components of
$\Theta^\m$ and $\bar\Theta^\m$.
One reason for believing this naive guess is that in a flat
background, where $h_{\m\nu}=\gamma^m_{\mu\nu}
(X_m -\bar X_m)+i (\gamma^n_{\mu \rho}\Theta^\rho)
(\gamma_{n \, \nu\sigma}\bar \Theta^{\sigma})$, the action reduces
to a free-field action that has been used successfully to calculate
S-matrix scattering amplitudes for the ten-dimensional superstring.$^{20}$
Furthermore, the physical vertex operators$^{21}$ that were used to calculate
these scattering amplitudes couple in the appropriate way to their
background fields in this sigma model action.

\vskip 24pt
\centerline {\bf Acknowledgements}
\vskip 12pt
I would like to thank Paul Howe for sharing with me his vast knowledge
on the subjects discussed in this paper, and for encouraging me to
work out the four-dimensional case in detail. I would also like to
thank Paul Townsend for telling me about the dilaton-coupling paradox,
George Papadopoulos for teaching me about N=(2,0) sigma models,
Joseph Buchbinder, Chris Hull, Christian Preitschopf, Anna Tollsten,
and Peter West for useful discussions,
and the SERC for its financial support.

\vskip 24pt

\centerline{\bf References}
\vskip 12pt

\item{(1)} Callan,C.G., Friedan,D., Martinec,E.J., and Perry,M.J.,
Nucl.Phys.B262 (1985), p.593.

\item{(2)} Mukhi,S., Nucl.Phys.B264 (1986), p.640.

\item{(3)} Witten,E., Nucl.Phys.B266 (1986), p.245.

\item{(4)} Carlip,S., Nucl.Phys.B284 (1987), p.365.

\item{(5)} Grisaru,M., Nishino,H., and Zanon,D., Nucl.Phys.B314
(1989), p.363.

\item{(6)} Gates,S.J.Jr., Majumdar,P., Oerter,R.N.,
and van de Ven,A.E., Nucl.Phys.B319 (1989), p.291.

\item{(7)} Gilbert,G. and Johnston,D., Phys.Lett.B205 (1988), p.273.

\item{(8)} Grisaru,M. and Zanon,D., Nucl.Phys.B310 (1988), p.57.

\item{(9)} Bergshoeff,E., Howe,P., Pope,C.N., Sezgin,E., and Sokatchev,E.,
Nucl.Phys.B354 (1991), p.113.

\item{(10)} Gates,S.J.Jr., Majumdar,P., Oerter,R.N.,
and van de Ven,A.E., Phys.Lett.B214 (1988), p.26.

\item{(11)} Townsend,P., private communication.

\item{(12)} Kraemmer,U. and Rebhan,A., Phys.Lett.B236 (1990), p.255.

\item{(13)} Gates,S.J.Jr. and Nishino,H., Class.Quant.Grav. 3 (1986), p.39.

\item{(14)} Berkovits,N., Phys.Lett.B232 (1989), p.184.

\item{(15)} Tonin,M., Phys.Lett.B266 (1991), p.312.

\item{(16)} Ivanov,E.A. and Kapustnikov,A.A., Phys.Lett.B267 (1991), p.175.

\item{(17)} Delduc,F., Galperin,A, Howe,P.,
and Sokatchev,E., ``A twistor formulation of the heterotic D=10
superstring with manifest (8,0) worldsheet supersymmetry", preprint
BONN-HE-92-19, JHU-TIPAC-920018, ENSLAPP-L-392-92, July, 1992.

\item{(18)} Sorokin,D.P., Tkach,V.I., Volkov,D.V., and Zheltukhin,A.A.,
Phys.Lett.B216 (1989), p.302.

\item{(19)} Berkovits,N., Nucl.Phys.B379 (1992), p.96., hep-th bulletin
board 9201004.

\item{(20)} Berkovits,N., ``Calculation of Greeen-Schwarz Superstring
Amplitudes using the N=2 Twistor-String Formalism'', SUNY at Stonybrook
preprint ITP-SB-92-42, August 1992, to appear in
Nucl.Phys.B, hep-th bulletin board 9208035.

\item{(21)} Berkovits,N., Phys.Lett.B300 (1993), p.53.

\item{(22)} Ogievetsky,V.I. and Sokatchev,E.S., Phys.Lett.79 (1978), p.222.

\item{(23)} Delduc,F. and Sokatchev,E., Class.Quant.Grav.9 (1992), p.361.

\item{(24)} Dine,M. and Seiberg,N., Phys.Lett.B180 (1986), p.364.

\item{(25)} Penrose,R. and MacCallum,M.A.H., Phys.Rep.6C (1972), p.241.

\item{(26)} Gates,S.J.Jr., Grisaru,M.T., Rocek,M., and Siegel,W.,
Superspace or 1001 Lessons in Supersymmetry, Benjamin/Cummings Publishing,
1983.

\item{(27)} Gates,S.J.Jr., Hull,C.M., and Rocek,M., Nucl.Phys.B248 (1984),
p.157.

\end